\documentclass[journal]{vgtc}                     

\onlineid{1420}

\vgtccategory{Research}

\vgtcpapertype{application/design study}

\usepackage{amssymb}
\usepackage{amsmath}
\usepackage{array}

\def\qcvis{Quantum Image Visualizer} 
\def\expert{E1} 

\def\overview{Image View}
\def\modalityView{Modality View}
\def\variationView{Variation View}

\def\detailview{Pixel View}
\def\probabilityView{Color Probability View}
\def\diffView{Difference Views}
\def\absdifferenceView{Absolute Difference View}
\def\differenceView{Difference View}

\usepackage[american]{babel}
\usepackage[pagebackref,bookmarks]{hyperref} 
\addto\extrasamerican{%
}

\newcommand{\Ket}[1]{\left|#1\right\rangle}

\title{\qcvis: Visual Debugging of Quantum Image Processing Circuits}

\author{%
  \authororcid{Anja Heim}{0000-0002-3670-5403},
  \authororcid{Thomas Lang}{0000-0001-5939-3919},
  \authororcid{Alexander Gall}{0000-0002-3649-7368},
  \authororcid{Eduard Gröller}{0000-0002-8569-4149},
  \authororcid{Christoph Heinzl}{0000-0002-3173-8871}
}

\authorfooter{
  \item
        Anja Heim and Thomas Lang are with the Fraunhofer Institute of Integrated Circuits IIS, Division Development Center X-ray Technology.
        E-mail: [anja.heim | thomas.lang]@iis.fraunhofer.de.
      \item
        Alexander Gall is with University of Passau, Faculty of Computer Science and Mathematics.
        E-mail: alexander.gall@uni-passau.de.
    \item
        Eduard Gröller is with TU Wien, Institute of Visual Computing \& Human-Centered Technology and VRVis Research Center.
        E-mail: groeller@cg.tuwien.ac.at.
  \item
        Christoph Heinzl is with University of Passau, Faculty of Computer Science and Mathematics and the Fraunhofer Institute of Integrated Circuits IIS, Division Development Center X-ray Technology.
        E-mail: christoph.heinzl@uni-passau.de.
}

\abstract{%
Quantum computing is an emerging field that utilizes the unique principles of quantum mechanics to offer significant advantages in algorithm execution over classical approaches. This potential is particularly promising in the domain of quantum image processing, which aims to manipulate all pixels simultaneously. However, the process of designing and verifying these algorithms remains a complex and error-prone task. To address this challenge, new methods are needed to support effective debugging of quantum circuits. The \qcvis{} is an interactive visual analysis tool that allows for the examination of quantum images and their transformation throughout quantum circuits. The framework incorporates two overview visualizations that trace image evolution across a sequence of gates based on the most probable outcomes. Interactive exploration allows users to focus on relevant gates, and select pixels of interest. Upon selection, detailed visualizations enable in-depth inspection of individual pixels and their probability distributions, revealing how specific gates influence the likelihood of pixel color values and the magnitude of these changes.
An evaluation of \qcvis{} was conducted through in-depth interviews with eight domain experts. The findings demonstrate the effectiveness and practical value of our approach in supporting visual debugging of quantum image processing circuits.

}


\graphicspath{{figs/}{figures/}{pictures/}{images/}{./}} 

\usepackage{mathptmx}                  

\begin{document}


\firstsection{Introduction}
\maketitle
Image processing of large datasets, such as those obtained in X-ray computed tomography (XCT)~\cite{Heinzl2017, Semmler2023}, is becoming an increasing challenge. Quantum computing offers a promising approach to apply image processing methods on quantum devices to achieve computational advantages~\cite{Yan2020}. Quantum image processing (QIMP) using the Novel Enhanced Quantum Representation (NEQR)~\cite{Zhang2013} for image encoding is particularly appealing as it allows to encode all pixels simultaneously in a superposition state. NEQR enables operations to be performed on all pixels of the image at once.

Since computation time on real quantum hardware is expensive, and since both reliability and performance are still limited, designers are forced to develop and test algorithms on simulators~\cite{Karafyllidis2005}. Designing QIMP algorithms is prone to errors when implementing gates for image preparation and processing, since the circuit sizes become extremely large even for small images. These challenges are further aggravated by the quantum nature of the images. Each pixel may be in a superposition state, thereby capable of assuming all possible colors within a specified bit depth. Consequently, each pixel may be characterized by a univariate probability distribution. Currently, QIMP algorithm designers must rely on manual recalculations or basic charts, such as the Bloch sphere~\cite{Bloch1946}, to identify errors due to the absence of advanced visualization tools.

To address these challenges, we present \qcvis{}, a novel tool for the analysis of QIMP algorithms for NEQR images. The main contributions of our design study are:
\begin{itemize}
    \item The identification of the design requirements for visual debugging of QIMP algorithms.
    \item The conception and implementation of an interactive visualization framework for analyzing and debugging QIMP circuits. \qcvis{} enables algorithm developers to explore the preparation and processing of quantum images step by step.  It provides a comprehensive view of gate operations on the entire image using two comparative visualization techniques. In addition, three different visualization techniques are proposed to examine and compare color probability distributions of individual pixels. 
    \item A detailed case study describing five different use cases showing how \qcvis{} is used to analyze correct and debug incorrect circuit implementations.
    \item A qualitative user study with seven QIMP developers working in the domain of XCT imaging to demonstrate the effectiveness and usability of \qcvis{}.
\end{itemize}

\section{Related Work}
Our work relates to the visualization of quantum circuits and the comparative visualization of images and data distributions.

\subsection{Visualization for Quantum Computing}
As outlined by Ruan et al.~\cite{Ruan2024}, visualization techniques for quantum computing typically focus either on representing individual quantum states or on illustrating quantum algorithms.
%
Quantum state visualizations are generally classified into two categories: vector-based approaches, which depict measured probability amplitudes, and probability-based approaches, which show computed probabilities. The Bloch sphere~\cite{Bloch1946} remains the most widely used method for visualizing quantum states. However, this visual metaphor can only encode a single quantum bit (i.e., qubit). Other geometric representations, e.g. as introduced by Altepeter et al.\cite{Altepeter2009} or Ruan et al.\cite{Ruan2023a} are similarly limited to visualize one- and two-qubit states. A geometry-based visualization for a few qubits is proposed by Galambos et al.~\cite{Galambos2012}.

%
When illustrating quantum algorithms, common methods involve the visualization of probabilities of the quantum states to represent the effects of individual operations in a quantum circuit.
%

While the current methods for quantum state and quantum algorithm visualization address various aspects of quantum computing, they are not well-suited for QIMP tasks. Many designs are limited to handling up to five qubits, making them impractical for the larger state spaces as required by NEQR images. Moreover, tools like QuantumEyes~\cite{Ruan2024} that track probability evolution are less useful for QIMP, where all quantum states, are typically modified simultaneously. This unique characteristic of QIMP thus demands visualization techniques tailored to image-based quantum algorithms.

\subsection{Comparative Visualization of Images and Distributions}
Comparative visualization enables users to assess differences between multiple data points or datasets and can be categorized into three main approaches: juxtaposition, superposition, and explicit encoding~\cite{Gleicher2018}. 
Image comparison is a crucial task in various domains and is often performed using either juxtaposed or superimposed representations. For instance, the VAICo system~\cite{Schmidt2013} facilitates hierarchical exploration of multiple images by highlighting local, small-scale variations. Scalable Insets~\cite{Lekschas2019} help users compare patterns in large images while minimizing navigation effort. Both approaches rely on superimposed representations after identifying differences to simplify user comparisons. Dynamic Volume Lines~\cite{Weissenbock2019} aggregates 3D XCT volumes into one-dimensional Hilbert Lines and functional boxplots~\cite{Sun2011}. This explicit encoding allows users to compare the aggregated XCT volumes. In \qcvis{}, we use explicit encoding and juxtaposition to compare NEQR images throughout their preparation and processing by quantum circuits. Explicit encoding highlights differences directly, eliminating the need for manual comparison. Juxtaposition ensures that users can analyze the NEQR images at any time without the cognitive burden of interpreting overlapping visual encodings.

Blumenschein et al.~\cite{Blumenschein2020a} offer a detailed review of common chart designs for distributions, including variations of line charts, histograms, and combinations thereof. The literature does not clearly indicate the optimal design for comparing distributions. We rely on the findings of Javed et al.~\cite{Javed2010} and Heim et al.~\cite{Heim2024} and use a combination of bar and line coding in a juxtaposed style. 
Comparing distributions is also a central task in time-series analysis. Line charts are typically compared using either superposition~\cite{Deng2024, Poco2014} or juxtaposition~\cite{Dasgupta2015, Lekschas2021}. Although juxtaposition requires users to shift their view across different areas of the screen, Javed et al.~\cite{Javed2010} found that juxtaposed arrangements outperform superposed designs for comparison tasks. Consequently, \qcvis{} uses a juxtaposed layout to visualize and compare the color probability distributions of NEQR images. 

\section{Background}
This section briefly explains the basics of QIMP. 

\subsection{Quantum Image Processing} 
QIMP enables parallel processing and exponential memory savings by encoding images in quantum states~\cite{Chauhan2020}. Especially in the domain of XCT, where images of considerable size are generated, this quantum advantage is currently under investigation. Domain experts are exploring different techniques for representing and processing XCT images on quantum devices~\cite{Semmler2023}. However, due to the current limitations of gate-based quantum computers, such as the small number of qubits and short decoherence times~\cite{Ruan2023a}, the processing of large images remains a challenge. Consequently, initial QIMP research in the domain of XCT focuses on the representation and processing of 2D grayscale images. Thus, our visualization framework focuses on representing quantum states describing 2D grayscale images.

In addition, given the high cost of testing on quantum hardware, classical simulators such as qiskit~\cite{JavadiAbhari2024} are employed to simulate quantum operations~\cite{Karafyllidis2005}. Currently, simulators are capable of handling larger images and more complex algorithms than real quantum hardware~\cite{Lang2025}. Simulators also provide much more detailed information, including the full quantum statevector (see \autoref{sec:neqrintro})~\cite{Semmler2023}, which is not collectible from quantum devices. For these reasons, domain experts mainly rely on simulators for algorithm development. The full quantum statevector, computed by the simulator, is utilized by our visualization framework for a comprehensive analysis.

\subsection{Quantum States and Quantum Image Representation}\label{sec:neqrintro}

A single quantum bit, i.e., qubit, can represent $0$ or $1$ as a classical bit. However, due to the superposition principle~\cite{von_neumann_mathematische_1932}, qubits exist in a state that can be regarded as having a probability of being $0$ and $1$. A single qubit $\Ket{\psi}$ can be described by 
$\Ket{\psi} = \alpha\Ket{0}+\beta\Ket{1},$
 where $\Ket{0},\Ket{1}\in\mathbb{R}^2$ are basis states and represent the classical $0$ and $1$. $\alpha,\beta\in\mathbb{C}$ are probability amplitudes for which applies 
$|\alpha|^2 + |\beta|^2 = 1$.
The squared magnitudes $|\alpha|^2$ and $|\beta|^2$ determine the probabilities of measuring the qubit in state $\Ket{0}$ or $\Ket{1}$, respectively. Upon measurement, the qubit collapses to one of these basis states based on the associated probabilities.

Multiple qubits may be combined into higher-dimensional states by means of the Kronecker product, i.e., given two single qubit states $\Ket{\psi_1}$ and $\Ket{\psi_2}$, a composed quantum state is formed via $\Ket{\psi}=\Ket{\psi_1}\otimes\Ket{\psi_2}$. The resulting statevector of two qubits is then described by $\Ket{\psi} = \alpha\Ket{00} + \beta\Ket{01} + \gamma\Ket{10} + \delta\Ket{11}$, subject to the condition $|\alpha|^2 + |\beta|^2 + |\gamma|^2 + |\delta|^2 = 1$, where $\alpha,\beta,\gamma,\delta$ are the probability amplitudes.

\begin{figure}[tb]
 \centering
 \includegraphics[width=\linewidth]{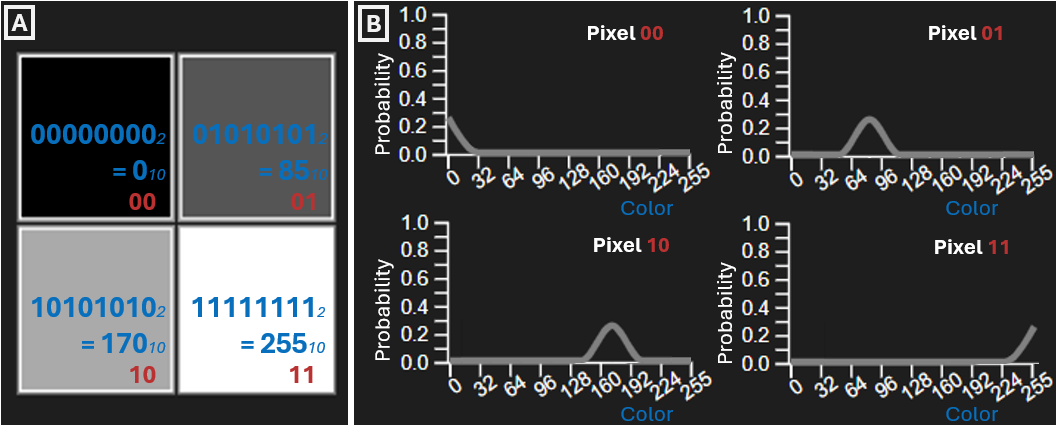}
 \caption{(A) Visual representation of the image encoded in Eq.~\eqref{eq:neqr_img_example}. (B) The probability distributions of the color values (blue) induced by the NEQR scheme for each pixel position (red) of the image.}
 \label{fig:background_neqr_explanation}
\end{figure}

The first step for image representation is to encode image pixels into quantum states. A widely used image encoding scheme is the NEQR~\cite{Zhang2013}, which allows all pixel values to be encoded and modified simultaneously.
Consider an image $I$ of size $2^n\times 2^n$ and $2^{b}$ grayscale color values in the range of $[0, 2^{b}-1]$, then NEQR encodes the pixel position information into $2n$ qubits and the grayscale color into $b$ qubits. The image is represented by the NEQR scheme as 
     $\Ket{I} = \Ket{XY}\otimes\Ket{C}$.
 The two states $\Ket{X}$ and $\Ket{Y}$ correspond to the image dimension in the x-direction, i.e., the image rows, and the y-direction, i.e., the image columns. The state $\Ket{C}$ represents the grayscale color. In NEQR the $2n$ position qubits are in superposition, such that the quantum state represents all the pixel positions at the same time. Next, the color value of a given pixel can be encoded. To illustrate this process, consider an $8$-bit $2 \times 2$ image, whose pixels are assigned to the following grayscale colors: $(0,0)\mapsto 0_{10} = 00000000_2$, $(0,1)\mapsto 85_{10} = 01010101_2$, $(1,0)\mapsto 170_{10} = 10101010_2$, $(1,1)\mapsto 255_{10} = 11111111_2$ (see \autoref{fig:background_neqr_explanation}(A)). The image can be represented as \cite{Yan2020}:
 \begin{equation} \label{eq:neqr_img_example}
    \begin{split}
        \Ket{I} = 2^{-1} (&\textcolor{Maroon}{\Ket{00}} \otimes \textcolor{NavyBlue}{\Ket{00000000}} + 
                   \textcolor{Maroon}{\Ket{01}} \otimes \textcolor{NavyBlue}{\Ket{01010101}} + \\
                  &\textcolor{Maroon}{\Ket{10}} \otimes \textcolor{NavyBlue}{\Ket{10101010}} + 
                   \textcolor{Maroon}{\Ket{11}} \otimes \textcolor{NavyBlue}{\Ket{11111111}})
    \end{split}
 \end{equation}
 
 In \autoref{eq:neqr_img_example}, the pixel position indices are displayed in red, while the color values are shown in blue. When encoding an image into quantum states using the NEQR scheme, the quantum representation includes not only the states corresponding to pixel positions and their associated colors, but also all possible linear combinations of these states, each with a specific probability. For example, even if the goal is to associate the pixel position $\Ket{00}$ exclusively with the color $\Ket{00000000}$ and assign this state the highest probability, the quantum nature of qubits would allow the pixel position $\Ket{00}$ to exhibit a probability distribution across all possible color states from $\Ket{00000000}$ to $\Ket{11111111}$. As a result, each pixel can simultaneously represent one, several, or all possible color values. The primary objective in NEQR encoding is to minimize the probability of these undesired color states and achieve a unimodal probability distribution, as illustrated by the line charts in \autoref{fig:background_neqr_explanation} (B). In essence, each pixel in a quantum image is described by a probability distribution that determines the probability of each possible color value occurring.

\subsection{Quantum Image Processing Algorithms}
Quantum images are prepared and processed using quantum circuits (see \autoref{fig:modality_view} (A)), where qubits are manipulated by applying quantum gates~\cite{Chauhan2020}. In NEQR this transformation affects the probability distributions of the color values of the pixel positions. To generate a specific image in the NEQR scheme, qubits are manipulated using a sequence of basic quantum gates, such as \textit{H} (Hadamard), \textit{X} (Not), and \textit{MCX} (Multi-Controlled Not) gates. In \autoref{fig:modality_view}(A), the \textit{NEQR} gate is a composite operation consisting of multiple gates applied to encode the desired pixel values. Since a large number of gates is required to set the color for each pixel, composite gates help manage circuit size and complexity for algorithm developers. Any subsequent image processing operation is also implemented by quantum gates. After applying the \textit{NEQR} gate in \autoref{fig:modality_view}(A), an \textit{X} gate is used on all qubits representing the gray values ($C_0$,...,$C_7$) to invert the color for each pixel. This is followed by another (incorrect) composite gate that should set the pixel at position $(0,0)$ to the color value $255$.

The development of QIMP circuits comes with a high risk of errors, which arise from using incorrect gates, implementation mistakes within gates, incorrect gate ordering, misunderstandings of framework conventions, etc. Currently, detecting such bugs requires manual inspection of gate implementations, step-by-step mathematical verification, or analyzing exponentially large statevectors, which is highly time consuming. 
We address this issue by the development of a visualization framework to aid algorithm developers in understanding and debugging QIMP algorithms. By providing visualization techniques for interpreting NEQR quantum statevectors, this framework aims to facilitate debugging workflows, reducing cognitive load, and saving time for developers of QIMP algorithms. 

\section{Design Principles}
Our design study addresses the requirements of developers of QIMP algorithms within the domain of XCT. Following the design study methodology by Sedlmair et al.~\cite{Sedlmair2012}, we collaborated with domain experts over a period of ten months in order to understand their workflows and challenges with the qiskit simulator~\cite{JavadiAbhari2024}. 

\subsection{Participants} \label{sec:participants} We collaborated with eight domain experts (1 female and 7 males, $age_{mean}=45$, $age_{sd}=10$). These domain experts integrated different educational backgrounds, i.e.,  one participant is a computer scientist (\expert{}), one is a mathematician (P1), and six of them are physicists (P2-P7), all working mainly in the field of QIMP for XCT imaging. Our experts also covered different levels of expertise (i.e., two of them are professors, five are postdoctoral researchers, and one postgraduate researcher) and have on average 3 years of experience in QIMP. In the development phase, we worked closely with \expert{} due to the expert's extensive practical experience of implementing QIMP algorithms. In the evaluation phase, all experts (P1-P7) except \expert{} participated. 

\subsection{Design Study Phases}\label{sec:phases} 
The design study consisted of three phases.

\textbf{In the initial phase}, we gathered information and established design requirements. 
During this phase, covering the first four weeks, we held weekly group meetings (1-2 hours) where each expert presented their previous work and workflows. The experts used the Bloch Sphere and qiskit histograms for visualization but relied primarily on manual calculations to verify gate executions due to the limitations of these tools. The Bloch Sphere~\cite{Bloch1946} visualizes single qubits only, while histograms show the probabilities of statevectors without detailing how gates affect the NEQR representation. This made understanding gate combinations and error detection challenging. After each presentation, participants shared future visualization requirements, which we documented and used to guide the development of our initial prototype.

\textbf{In the development phase}, we refined our prototype based on expert feedback. In this phase, which covered over nine months, we iteratively refined our mock-ups and our prototype with expert \expert{} based on the qualitative requirements from the initial phase. Weekly, open-ended meetings allowed \expert{} to provide feedback that guided adjustments to the visual design to ensure that the prototype met real-world needs.

\textbf{In the evaluation phase}, we tested our visualization framework to assess its effectiveness and usability (see \autoref{sec:eval}).

\subsection{Design Requirements}
The overall goal of \qcvis{} is to provide a clear and comprehensive visualization of the QIMP pipeline to enhance the understanding of the algorithm's structure and execution. Such an overview may also serve as educational tool to explain quantum imaging concepts and workflows. The resulting four design requirements from the design study process are defined as follows:

\textbf{R1 - Visual Representation of NEQR.}
Enable the visualization of the NEQR scheme to display the statevectors as images, facilitating error identification during image preparation. Accurate pixel encoding is essential before analyzing the subsequent steps of the algorithm.

\textbf{R2 - Visualize Gate Effects on the Image.}
Visualize the overall effect of quantum gates on the entire image, allowing users to identify which gates have the most significant effect on the image or specific image regions.

\textbf{R3 - Visualize Gate Effects on the Individual Pixels.}
Provide a detailed view of how quantum gates influence individual pixels, particularly regarding their color probability distributions, to support in-depth analysis and debugging.

\textbf{R4 - Scalability.}
The visualization framework should be designed to accommodate images encoded with a large number of qubits and algorithms consisting of many gates. Thus, it should be usable for image processing algorithms of varying complexity. 

The smallest image representable via NEQR is a binary ($b=1$) image of size $2 \times 2$ ($n=1$) requiring $3$ qubits for implementation. The maximum computable size depends on the available RAM of the workstation. Our domain experts typically work with image sizes of $4 \times 4$ or $8 \times 8$~\cite{Semmler2023}. Larger images demand more memory and computational power, significantly increasing computation time. The largest image successfully tested by our experts was of size $16 \times 16$.
%
The number of gates grows with the number of qubits and the complexity of the operations. Depending on the image size, hundreds of basic gates, i.e., \textit{X} or \textit{MCX} gates, may be required for image preparation alone. Therefore, domain experts rely on composite gates to combine the most important gate operations and reduce the circuit size for a better overview in the circuit diagram. However, the number of composite gates required for image preparation and processing can quickly grow from tens to hundreds of gates as the image size increases~\cite{Semmler2023}.

\section{\qcvis} \label{sec:methodology}
\qcvis{} is an interactive visualization framework for the analysis of quantum images represented using the NEQR scheme and processed by quantum circuits. To explain the framework, we use an 8-bit (256 grayscale values) image of size $2 \times 2$ (see \autoref{fig:background_neqr_explanation}(A)) prepared and processed by a circuit with three gates (see \autoref{fig:modality_view}(A)). Our exemplary circuit to demonstrate \qcvis{} includes: (1) a composite \textit{NEQR} gate that sets the grayscale values for all pixels, (2) an \textit{X} gate applied to the color qubits, and (3) a composite gate intended to set pixel (0,0) to the value white (255).

\textbf{Implementation.}
\qcvis{} is implemented using Python 3.12, and PySide6~\cite{PySide6}, with all visualization techniques built using D3.js~\cite{Bostock2011}. It is designed to visualize and debug quantum circuits created with qiskit~1.2~\cite{JavadiAbhari2024}.

\textbf{Data Preprocessing.}
For visualization, the gates and the statevectors of the qubits are passed to \qcvis{} (see \autoref{fig:teaser} (left)). For each gate operation, a list of statevectors is obtained representing the image. These vectors provide a probability for all possible grayscale color values at each pixel position. For the example circuit, a statevector of $2^{\textcolor{Maroon}{2}+\textcolor{NavyBlue}{8}} = 1024$ elements is received for each of the three gates.
During preprocessing, the statevectors are decomposed into pixel positions, grayscale color values, and their associated probabilities. Each image is described by its pixels, and each pixel contains a probability distribution over all possible grayscale color values (see \autoref{fig:background_neqr_explanation}(B)). This processed data forms the basis for the visualization.

\textbf{Views.}
\qcvis{} provides users a circuit diagram, a general overview of the NEQR image in the \overview{}, and a detailed inspection of the pixels' color probability distributions in the \detailview{}. 
%
The circuit diagram is equivalent to that generated by the simulator (see \autoref{fig:modality_view} (A)). It provides experts with a familiar overview of qubits, gates, and their interactions. Moreover, it facilitates the comprehension of the algorithm and the validation of insights derived from the \overview{} and \detailview{}.

\subsection{\overview{}}\label{sec:imageView}
\begin{figure}[b]
  \centering
  \includegraphics[width=\linewidth]{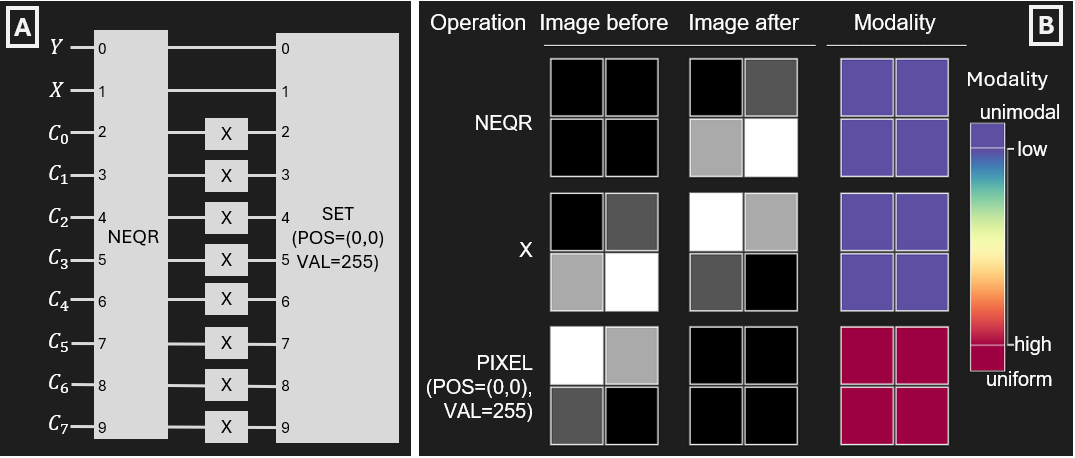}
  \caption{(A) Circuit preparing and processing the image in \autoref{fig:background_neqr_explanation}. (B) \overview{}: Gate labels on the left, followed by the image representation before and after the gate was applied, and finally the \modalityView{}.
    }
    \label{fig:modality_view}
\end{figure}

This view represents the statevectors depicted as images, with gate labels displayed vertically from top to bottom. For each gate, the images before and after its application are shown. Each pixel is depicted as a square, colored according to the most probable value. In cases where multiple colors share the highest probability, the smallest (darkest) color value is used. As illustrated in \autoref{fig:modality_view}(B), the initial state shows all pixels as black (0). After applying the \textit{NEQR} gate, pixel (0,0) remains black, while pixels (0,1), (1,0), and (1,1) change to dark gray (85), light gray (170), and white (255), respectively. The subsequent gate inverts the colors by swapping each pixel’s color value with its binary opposite: black (0) becomes white (255), dark gray (85) becomes light gray (170), light gray (170) becomes dark gray (85), and white (255) becomes black (0).
The final gate is intended to set pixel (0,0) to white (255), but instead, all pixels incorrectly display black (0), suggesting a potential error in the gate. The \modalityView{} provides additional insights to further investigate this issue.

\subsubsection{\modalityView{}} \label{sec:modalityView}
The \modalityView{} visualizes the shape of the distribution for each pixel in the image.
%
When preparing an image, the goal is to maximize the probability of a specific color for each pixel. This results in a unimodal probability distribution with a single peak at the target color. A correctly implemented \textit{NEQR} gate produces this unimodal distribution for a pixel. Errors in this gate can cause multiple peaks. The worst case is a uniform distribution, where all color values have equal probability, making it impossible to reliably measure the intended pixel color.
%
The \modalityView{} uses color coding to provide an overview of the shape of these distributions. It combines a continuous and discrete color scheme based on the Spectral~\cite{Reda2018} colormap. 
The unimodal and uniform cases have specific colors, purple and red, respectively. These are the extreme cases, and are the most important and common ones. The more peaks there are in the distributions, the more the color changes from blue to orange.
The modality is computed by identifying and counting local maxima in the distribution~\cite{2020NumPy}. The number of peaks found is used for color coding. A single peak corresponds to a unimodal distribution, no peaks indicate a uniform distribution, and multiple peaks are mapped to the continuous color gradient.
In \autoref{fig:modality_view}(B), correctly implemented gates produce purple squares in the \modalityView{}, as shown for the \textit{NEQR} and \textit{X} gates. The last gate results in uniform distributions for all pixels, represented by red squares, indicating a significant error.
In advanced QIMP algorithms, temporary multimodal distributions may occur during computation. However, the final image should always return to a unimodal state for correct pixel measurements. If the \modalityView{} displays non-unimodal distributions at the end, the algorithm is likely incorrect.

\subsubsection{\variationView{}}\label{sec:variationView}
\begin{figure}[tb]
  \centering
  \includegraphics[width=\linewidth]{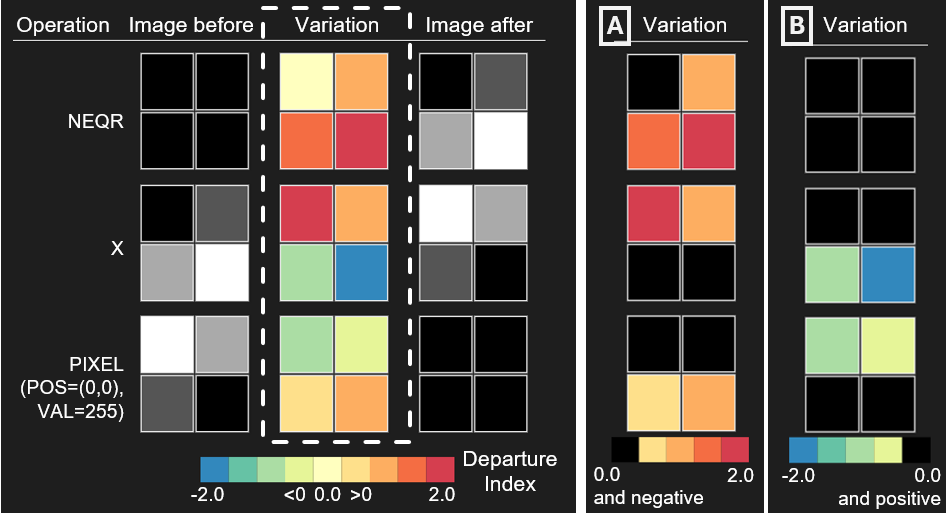}
  \caption{\variationView{} positioned in between the grayscale images. It shows the departure of the distributions for each pixel. (A) and (B) show filtering through user interaction.
  }
    \label{fig:variation_view}
\end{figure}

The \variationView{} provides insight into how gate operations modify pixels' color probability distributions, even if the modality remains unchanged. It visualizes these changes by quantifying the magnitude and direction of the distribution shift between the image before and after a gate is applied (see \autoref{fig:variation_view}).
%
A positive shift indicates that the peak of the distribution is moving toward a brighter color (e.g., from dark gray to light gray), while a negative shift represents a transition to a darker color. This information allows users to assess whether and to what extent a gate is brightening or darkening specific pixels.
%
The \variationView{} uses a color-coded image positioned between the images before and after the gate, facilitating direct comparison. It applies a discrete, nine-color scheme from the Spectral colormap~\cite{Reda2018} to visualize these shifts based on the departure index~\cite{Menning2007}. This index measures the difference between two distributions.
The departure index $M$ quantifies the difference between a test distribution and a reference distribution, providing both the direction and magnitude of the shift along the horizontal axis. If $M > 0$, the test distribution is shifted positively, i.e., to the right and towards brighter colors, and if $M < 0$, it is shifted negatively, i.e., to the left and towards darker colors. If $M = 0$, no shift and thus no difference is detected between the two distributions.
The magnitude of the departure index indicates the extent of this shift, ranging between $-2$ and $2$.
In the \variationView{}, this shift is visualized using a discrete color scheme. A negative shift is mapped in the color scheme to four discrete colors from light green to blue with decreasing magnitude. A positive shift is mapped from light orange to red with increasing magnitude. No shift is shown in yellow.
Each pixel is color-coded based on its departure index, allowing users to quickly identify whether the applied gate alters the pixel values as expected.
The departure index was chosen as it not only quantifies differences but also indicates direction and magnitude, information which is not provided by other measures like the Kolmogorov-Smirnov test~\cite{Berger2014} or the Kullback-Leibler divergence~\cite{Song2002}.
In \autoref{fig:variation_view}, the \textit{NEQR} gate generates positive shifts for all pixels except (0,0), which remains unchanged (yellow). Increasing magnitudes of positive shifts are shown by a gradient from light orange to red. The \textit{X} gate performs a color inversion, whereby dark pixels (0,0) and (0,1) become brighter (orange and red), while bright pixels (1,0) and (1,1) become darker (green and blue). Although the \modalityView{} reveals an error in the final gate, the \variationView{} demonstrates how it alters the distributions, transforming them from strongly skewed to uniform. This transformation results in a moderate negative shift (greenish) for pixels (0,0) and (0,1), and a moderate positive shift (orangy) for the others (see distributions in \autoref{fig:probability_view}).
The \variationView{} complements the grayscale images by highlighting pixel-wise changes that might be imperceptible through grayscale contrast alone. It is especially useful for detecting errors where incorrect pixel values are computed or unintended pixels are modified.

\subsubsection{User Interaction}
In \qcvis{}, users can hide gates by clicking on them (i.e., the \textit{hide} interaction). This reduces the number of gates to inspect and allows the user to perform a detailed analysis of important gates.

The grayscale image representation supports two types of user interactions.
First, users can hover over individual pixels to display a tooltip that shows the exact color value and its corresponding probability. This interaction helps users accurately examine pixel colors, especially when subtle differences in gray levels are hard to distinguish.
%
Second, users can select specific pixels by clicking on the "before" or "after" images. This action opens a larger view of the selected image, making pixel selection easier. The system supports both brushing and multi pixel selection. The selected pixels appear in the \detailview{} when the window is closed and are highlighted with a pink border in the image representations.

The \variationView{} allows users to filter the color scheme, focusing on either positive or negative shifts (see \autoref{fig:variation_view}(A) and (B)). Filters may be applied by hovering over the color scheme and using the scroll wheel to switch between the original view and the filtered views. This helps to identify gates that have either a positive or negative effect on pixel values. Pixels outside the selected range are displayed in black.

\subsection{\detailview{}}\label{sec:pixelView}
The \overview{} provides an overview of the quantum images, providing only general indicators regarding the color probability distributions of the pixels. In contrast, the \detailview{} allows users to explore these distributions in depth, enabling a clear understanding of how individual gates affect the color probabilities. This view supports three different visualization techniques, each of which presents the distributions in a different context. The \probabilityView{} provides detailed insights into the distribution itself, while the other two techniques visualize the differences between the distributions.

\subsubsection{\probabilityView{}} \label{sec:probability_view}
\begin{figure}[htb]
  \centering
  \includegraphics[width=\linewidth]{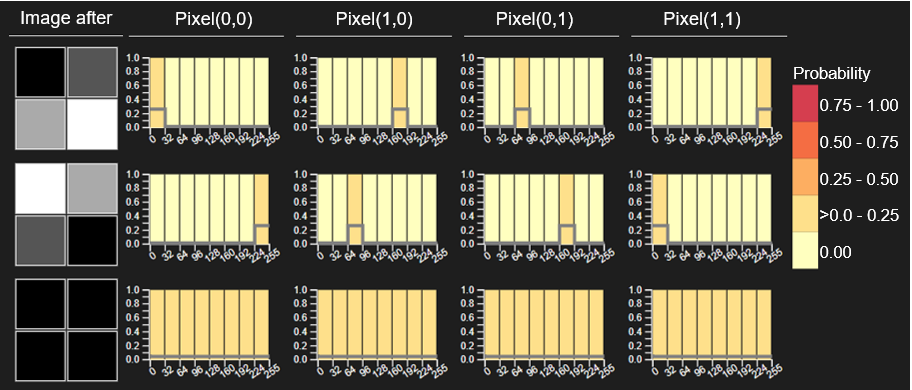}
  \caption{\probabilityView{} visualizing the color probability distribution accumulated in eight bins of all four selected pixels.
    }
    \label{fig:probability_view}
\end{figure}

The \probabilityView{} visualizes the color probability distribution for each selected pixel over all gate operations.
%
For each selected pixel in the \overview{}, a histogram of its distribution is computed. Each row of the \detailview{} grid represents an executed gate, while each column corresponds to a selected pixel. Within each cell, the binned distribution for a specific pixel is displayed. This visualization combines both bar and line encodings inspired by the AccuStripes representation~\cite{Heim2024}. Probability values are represented by the color of the bars in the background and by the height of the line, which in combination improves legibility. The horizontal axis shows the binned color ranges, and the vertical axis represents the probability values, ranging from $0.0$ to $1.0$. For color coding, a discrete color scheme derived from Spectral~\cite{Reda2018} is used, where higher probabilities transition from light yellow to red.
%
\autoref{fig:probability_view} illustrates the distributions for each pixel of the example image. For instance, pixel (0,0) initially shows the highest probability in the leftmost bin (color range 0-32). After the second gate, the peak shifts to the rightmost bin (color range 224-255), while the final gate changes the distribution to a uniform one. This combined bar-line encoding allows for precise examination of distributions, even when they contain small but significant peaks.

\subsubsection{\diffView}
\begin{figure}[tb]
  \centering
  \begin{subfigure}[b]{\columnwidth}
  	\centering
  	\includegraphics[width=\textwidth]{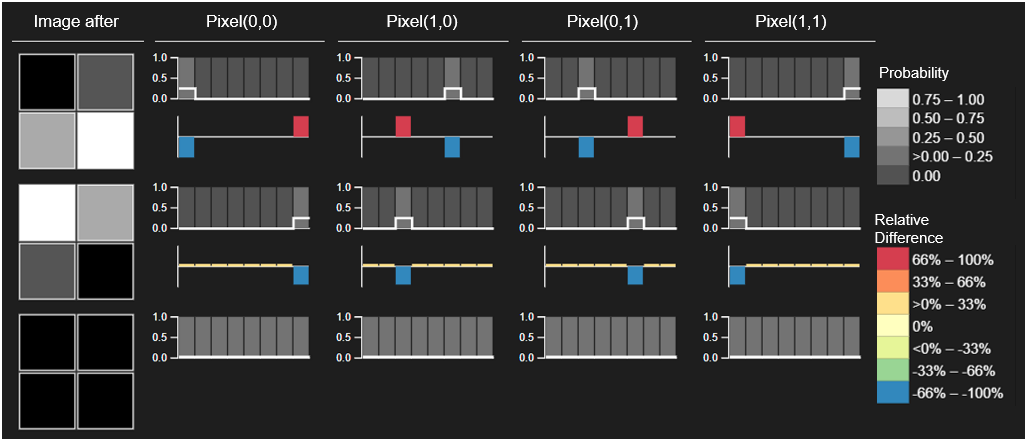}
  	\caption{\differenceView{} showing the relative difference per bin of the distributions pairwise from top to bottom.}
  	\label{fig:rel_diff_view}
  \end{subfigure}%
  \caption{\diffView{} providing two possibilities for to support the inspection of the pairwise differences in the distributions, absolute and relative.}
  \label{fig:pixel_view}
\end{figure}

In addition to examining the structure of color probability distributions, it is equally important to understand how these distributions change between gate operations. To address this need, the \absdifferenceView{} (see \autoref{fig:pixel_view}(A)) and \differenceView{} (see \autoref{fig:pixel_view}(B)) focus on highlighting the differences between successive distributions.

In these views, each grid cell displays, the visualization of the distribution itself, as well as a bar chart, which highlights the difference between the distributions. The original distributions remain visible as a reference using the combined bar and line encoding. Compared to the \probabilityView{}, the visualization of the distribution occupies half of the space and employs a discrete grayscale color scheme ranging from dark to light gray in five steps. This design prioritizes the visualization of differences while maintaining the distribution for context.
The difference calculation involves normalizing the probability of each bin by the maximum probability within the corresponding histogram. Then, normalized values of matching bins in consecutive distributions are subtracted from each other. In the \absdifferenceView{}, the absolute differences are used for encoding, while the \differenceView{} retains the signed differences to indicate directionality.


The \differenceView{} displays the signed difference between distributions using a diverging bar chart. Bars representing positive differences extend upward from the center axis, while negative differences extend downward. The bar size is halved compared to the \absdifferenceView{} to accommodate this dual-direction encoding. A seven-step discrete color scheme from Spectral~\cite{Reda2018} is used, with positive differences ranging from yellow to red, and negative differences ranging from light green to blue. This view not only highlights which bins changed, but also whether the probabilities increased or decreased. 
\autoref{fig:pixel_view}(B) shows the signed differences for pixel (0,0) in the example image. For pixel (0,0), the first bin shows a decrease, while the last bin exhibits an increase. Between the second and third gate, probabilities increase slightly across all bins except the last, where a stronger decrease occurs.

Each visualization has distinct advantages. The \differenceView{} provides a more detailed view of how probabilities increase or decrease for each bin, but the smaller bar size may cause subtle differences to be overlooked during quick inspection. In contrast, the larger bars in the \absdifferenceView{} are more visually prominent, making it easier for users to spot significant changes across multiple gates and pixels.


\subsubsection{User Interaction}
Users can seamlessly switch between the three visualization techniques, similar to the interaction in the \variationView{}. By hovering over the color scheme and using the scroll wheel, users can browse through the different views. This interaction allows for quick and easy examination of specific pixels at various gates.
%
When users hover over a bin in any visualization that displays color probabilities, a tooltip appears. This tooltip provides detailed information about the color range contained in the bin and its current probability value (see \autoref{fig:UseCase1} (B)).

\subsection{Design Considerations}

\textbf{Color Scheme.}
%
\qcvis{} employs six distinct color schemes across its visualization techniques. 
Despite extensive research on optimizing individual color schemes, there is limited work addressing how to effectively integrate multiple color schemes within a visualization framework. Given the challenge of combining six unique color schemes without overlapping colors, we opted in collaboration with \expert{} to use two main palettes: variations of the Spectral color scheme~\cite{Reda2018} and a sequential grayscale~\cite{Brewer2003}.
%
The number of color steps was determined through expert testing. We used five distinct colors for sequential schemes and seven to nine for diverging schemes, ensuring a balance between precision and ease of interpretation. The \variationView{} uses nine colors to capture subtle differences, while the \differenceView{} uses seven colors, as finer distinctions were considered less critical.

\textbf{Design Alternatives.}
%
We explored several design alternatives for our visualization techniques. For the \variationView{}, we prioritized a simple design to facilitate quick comprehension. Within the \detailview{}, we considered line charts and histograms as alternatives for representing distributions. However, due to the potential for extreme probability variations, ranging from high to very low peaks, we decided against purely height-based encodings, as they could become difficult to read. We adopted a hybrid approach combining colored bars and line encodings, similar to AccuStripes~\cite{Heim2024}. This improves comparability of distributions across multiple gates.
The decision to use eight bins for the distribution was based on subjective testing with \expert{}. Since the distributions in our use cases did not exhibit excessive complexity, eight bins provided a balance between simplicity and detail.

\textbf{Layout.}
\qcvis{} supports two different layout configurations: vertical and horizontal. In the vertical layout, the circuit diagram appears at the top, followed by the \overview{} on the left and the \detailview{} on the right. In the horizontal layout, the circuit is displayed at the top, with the \overview{} below and the \detailview{} at the bottom. All figures in \autoref{sec:methodology} follow the vertical layout.
The vertical layout is tailored for comparing distributions, aligning with the experts’ familiarity with line charts and histograms, where values are represented vertically. The horizontal layout aligns with the reading direction of quantum circuits, which progresses from left to right. This layout required specific design adjustments for the \detailview{}. To maintain consistency, the color probability distributions of each pixel must appear directly below the gate that generates them. Consequently, the x-axis is positioned vertically (from bottom to top), while the y-axis runs horizontally (from left to right). This arrangement preserves the visual consistency of the \diffView{} in both layouts. 

\section{Case Study}\label{sec:casestudy}
We demonstrate the utility of \qcvis{} through a case study conducted with one of our domain experts, \expert{} (also a co-author).
%
The case study is divided into two main tasks: NEQR image preparation and processing. In the first task, \expert{} analyzed two image preparation scenarios, one correct and one incorrect. In the second task, \expert{} examined three processing algorithms: a correct and an incorrect color inversion, and a thresholding operation. The incorrect versions were included to evaluate how \expert{} would use \qcvis{} for error detection and analysis.
%
Since \expert{} was already familiar with the visual design and interaction techniques from prior collaborations, no additional training was required. To simulate a realistic debugging scenario, the use cases were implemented by one of the authors, ensuring that \expert{} was unaware of the specific bugs. \expert{} was informed about the number of scenarios, but not their specific objectives, such as preparing an image or applying a threshold operation. This approach allowed us to observe whether and how \expert{} identified errors using \qcvis{}. As \expert{} preferred the vertical layout, all figures in this section follow this orientation.

\subsection{Case Study I - NEQR Image Preparation} \label{sec:UseCase1}
In both scenarios, $8 \times 8$ images with 8-bit color depth (256 gray values) were prepared using 14 qubits and 65 gates. The circuit started with an \textit{H} gate, followed by the sequential application of pixel setter gates. Each pixel setting is named to indicate the pixel position and the corresponding gray value. For example, the gate \textit{PIXEL(POS=(6,5), VAL=238)} sets the pixel at row 6, column 5 to the gray value 238.

\begin{figure}[tb]
  \centering
  \includegraphics[width=\linewidth]{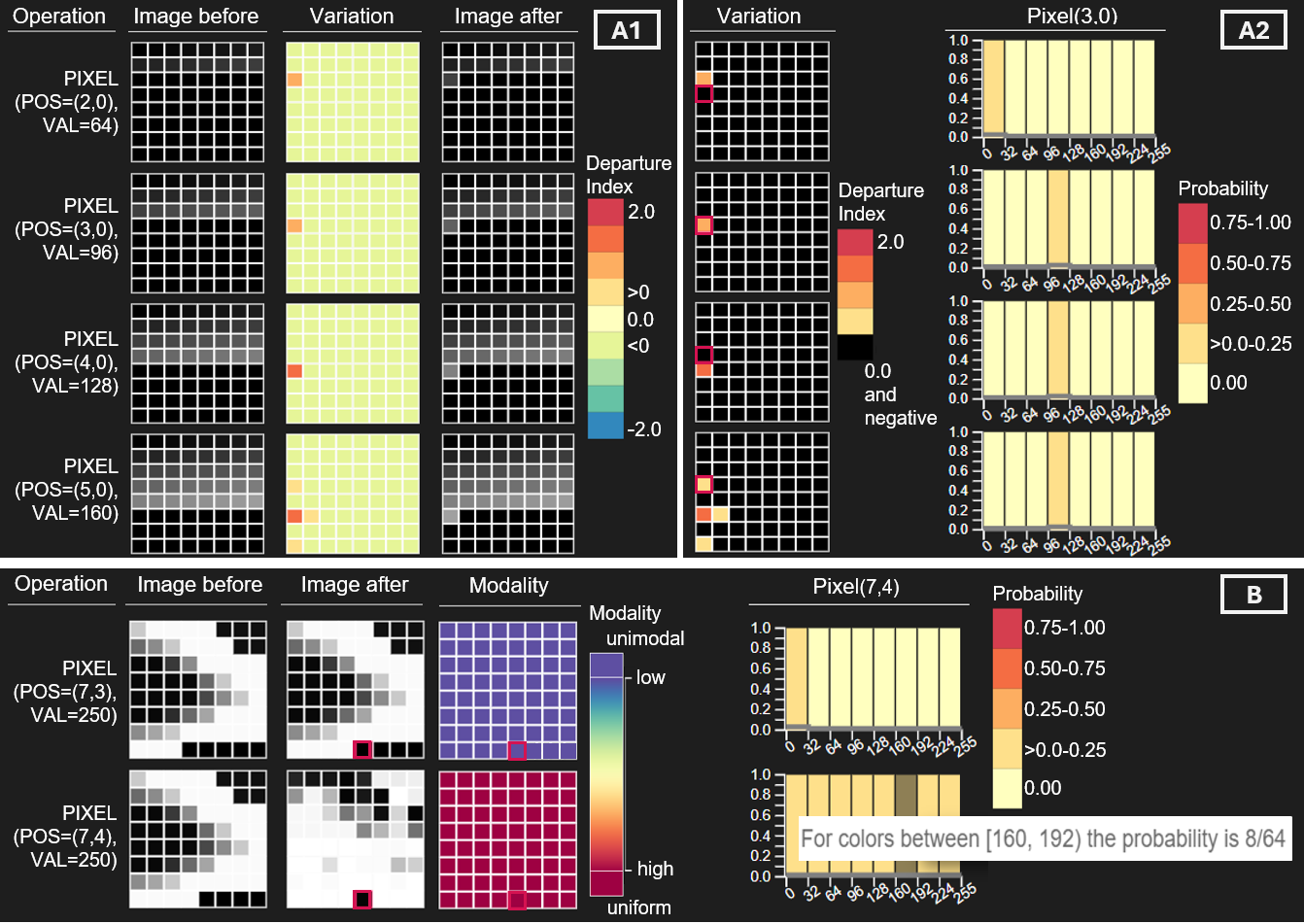}
  \caption{NEQR Image Preparation: (A1,A2) Correct and (B) Incorrect. (A1) Setter gates for pixels in column 0 are shown. (A2) Numerical instabilities are visible in the last gate. (B) Error in the second gate results in uniform color probability distributions for the pixels.}
    \label{fig:UseCase1}
\end{figure}

\textbf{Correct NEQR Preparation:} 
\expert{} began analyzing the circuit, observing the grayscale images (\textbf{R1}) as they evolved in the vertical layout. \expert{} identified that the pixel values were set sequentially, row by row, from left to right. Furthermore, \expert{} observed that all pixels in the \modalityView{} were purple, indicating a unimodal color distribution of each pixel.
%
Switching to the \variationView{} (\textbf{R2}), \expert{} examined the departure index and its color-coded representation (see \autoref{fig:UseCase1}(A1)). \expert{} used the \textit{hide} interaction to focus on the gates that are preparing the pixels in column 0, and removed all others from the view (\textbf{R4}).
%
By comparing the visualizations in the \variationView{}, \expert{} noticed that the pixel color increased gradually from orange to red. This led \expert{} to conclude that the gates cause a positive shift in the pixels' color probability distributions.

While inspecting the pixel setter gates for column 0, \expert{} identified unexpected behavior in the gate \textit{PIXEL(POS=(5,0), VAL=160)}, which altered multiple pixels instead of one. To clarify this anomaly, \expert{} adjusted the color scheme of the \variationView{} by hovering over the color legend and using the mouse wheel to change it (see \autoref{fig:UseCase1}(A2)). This adjustment revealed that, beyond the intended pixel (5,0), three additional pixels (3,0), (7,0), and (5,1) were also affected.
\expert{} selected pixel (3,0) for a detailed inspection in the \probabilityView{} (see \autoref{fig:UseCase1}(A2)). Tracing its distribution across multiple gates, \expert{} observed that gate "PIXEL(POS=(3,0), VAL=96)" shifted the probability peak from the first bin (colors 0-31) to the fourth bin (colors 96-127). Although the \variationView{} indicated changes in the distribution at gate \textit{PIXEL(POS=(5,0), VAL=160)}, there was no visible change in the \probabilityView{}.
This discrepancy suggested the presence of numerical instabilities in the simulator: the change was significant enough to be detected in the \variationView{} but too small to be reflected in the \probabilityView{}. This unexpected discovery, revealed new insights into how the simulator handles image preparation.


\textbf{Incorrect NEQR Preparation:}
In another circuit, an additional \textit{H} gate was introduced into the pixel setter gate \textit{PIXEL(POS=(7,4), VAL=250)}. Since only the gate name, which is defined by the algorithm developer, is visible, these implementation changes remain hidden within the circuit diagram. 
This scenario reflects real-world challenges where algorithm developers may correctly label composite gates according to their intended function but introduce errors when assembling the underlying basic gates.
%
Using \qcvis{}, \expert{} explored the circuit by inspecting the \overview{}. Scrolling through the images, \expert{} noticed that the pixel values were set sequentially, row by row, from left to right. The \modalityView{} (\textbf{R2}) consistently displayed purple colors for the pixels until the gate \textit{PIXEL(POS=(7,4), VAL=250)} (see \autoref{fig:UseCase1} (B)). At this gate, the \modalityView{} displayed red squares, indicating a uniform distribution for each pixel. Upon closer examination of the image resulting from this gate (see \autoref{fig:UseCase1} (B) (Image after)), \expert{} noticed that most pixels' color values had changed significantly compared to the previous image (see \autoref{fig:UseCase1} (B) (Image before)). \expert{} selected the pixel (7,4) and used tooltips in the \probabilityView{} (\textbf{R3}), that appeared when hovering over the bins, to inspect the exact probability values. \expert{} confirmed a shift from a unimodal distribution to a uniform distribution, where each bin had a probability of $\frac{8}{64}$.
Guided by the insights of \qcvis{}, \expert{} reviewed the qiskit implementation of gate \textit{PIXEL(POS=(7,4), VAL=250)} and identified the incorrect \textit{H} gate as disrupting the pixel structure within the quantum state.


\section{Qualitative User Study}\label{sec:eval}
In addition to the case study, we conducted an in-depth user interview with seven domain experts (P1-P7, see \autoref{sec:participants}) to assess the effectiveness and usefulness of \qcvis{}. 

\subsection{Study Design}


\textbf{Datasets.}
We used two distinct QIMP circuits for the evaluation, both derived from and extending the use cases presented in \autoref{sec:casestudy}. Circuit 1 (C1) encoded a 4 $\times$ 4 NEQR image with a color depth of 256 gray values. It comprised 20 gates, including 16 pixel setter gates, two \textit{X} gates, and two erroneous pixel setter gates. 
Circuit 2 (C2) processed a 4 $\times$ 4 image with a reduced color depth of eight gray values. It consisted of 18 gates, including 16 pixel setter gates, one threshold gate, and one \textit{X} gate.


\textbf{Tasks.}
We designed five tasks based on expert feedback gathered during the initial phase of our design study. These tasks address the most critical challenges in QIMP algorithm design. Tasks (T1-T3) focus on evaluating the effectiveness of \qcvis{} in providing a global analysis of NEQR images. 
Participants were asked to (T1) identify the gates with the most significant impact on pixels' colors, (T2) describe the overall form of pixel distributions, and (T3) explain roughly how the gates influence these distributions. 
The final two tasks (T4-T5) evaluate the framework's ability to support a detailed examination of individual pixel color probabilities.  
Participants were required to (T4) identify alternative colors likely to be measured for specific pixels and (T5) explain how the gates modify these color probabilities in detail.



\textbf{Procedure.}
We performed one-on-one, semi-structured interviews with seven experts (P1-P7). First, participants received a 15-minute tutorial on the visual design of all views in both horizontal and vertical layouts of \qcvis{}, using the image from \autoref{sec:methodology} (see \autoref{fig:modality_view}). Next, each participant completed the five predefined tasks using \qcvis{}, with the presentation order of the layouts counterbalanced across participants to reduce order effects. Each participant performed each condition exactly once. As they progressed through the circuit, participants were asked to verbally describe how the image evolved in the circuit. The average duration of the experiment was approximately 45 minutes. Following task completion, participants engaged in a 15-minute post-study interview, providing feedback on the visual designs. We took detailed notes throughout the experiment. Finally, participants rated \qcvis{} using a 5-point Likert scale questionnaire (see \autoref{tab:questionaire}).

\begin{table}[tbh]
    \centering
    \begin{tabular}{p{0.05\linewidth} | p{0.85\linewidth}}
        \hline
        Q1 & How does \qcvis{} help explain the process of QIMP? \\
        Q2 & To what extent does \qcvis{} support the analysis of the evolution of images in the quantum circuit? \\
        Q3 & To what extent does the \overview{} help in understanding the impact of gates on images? \\
        Q4 & How does the \detailview{} help in understanding pixel colors?\\
        Q5 &  To what extent do the \modalityView{} and the \variationView{} support debugging QIMP algorithms?\\
        Q6 &  To what extent do the \probabilityView{} and the \diffView{} support debugging QIMP algorithms?\\
        \hline
        Q7 & How would you rate the learning difficulty of \qcvis{}? \\
        Q8 & How useful is \qcvis{}? \\
        Q9 & Would you be interested in using \qcvis{} in the future to analyze QIMP algorithms? \\
        \hline
        Q10 & How would you rate the smoothness of user interaction? \\
        Q11 & How would you rate the ease of use of \qcvis{}? \\
        \hline
        Q12 & How would you rate the clarity of the overall design? \\ 
        \hline
        Q13 &  To what extent does the \textbf{vertical layout} support understanding QIMP algorithms for the \overview{}? \\
        Q14 &  To what extent does the \textbf{horizontal layout} support understanding QIMP algorithms for the \overview{}?\\
        Q15 &  To what extent does the \textbf{vertical layout} support understanding QIMP algorithms for the \detailview{}? \\
        Q16 &  To what extent does the \textbf{horizontal layout} support understanding QIMP algorithms for the \detailview{}? \\
        \hline
    \end{tabular}
    \caption{The questionnaire consists of five parts: the effectiveness for analyzing QIMP algorithms(Q1-Q6), the usability (Q7-Q9), the user interactions (Q10-Q11), the visual design (Q12), and the user preference for a specific layout (Q13-Q16).}
    \label{tab:questionaire}
\end{table}

\subsection{Results}
\begin{figure}[tb]
  \centering
  \includegraphics[width=\linewidth]{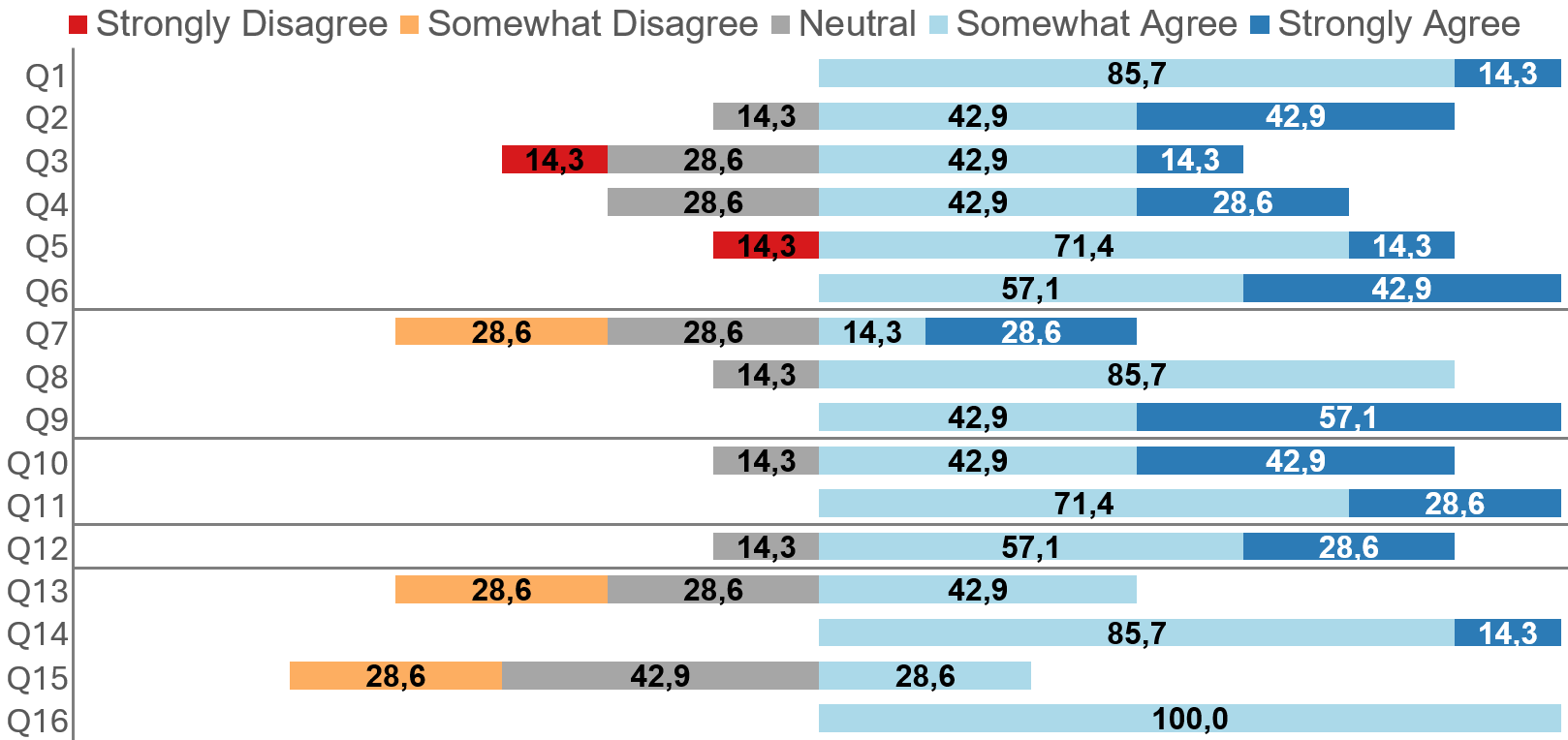}
  \caption{User feedback for the questionnaire in \autoref{tab:questionaire}.
    }
    \label{fig:likert_scale}
\end{figure}

\autoref{fig:likert_scale} summarizes the feedback collected from the questionnaire. We discuss the details in the following.

\textbf{Effectiveness for Analyzing and Debugging QIMP algorithms.}
The majority of participants agreed that \qcvis{} effectively supports the understanding and debugging of preparing and processing NEQR images. Participants P1-P4 particularly valued the integration of multiple visualization techniques. As P1 noted, "\textit{I can follow the circuit and how the gates affect the image step by step.}" P3 suggested that \qcvis{} could also serve educational purposes, stating, "\textit{Students could use the system to learn more about the topic.}" 
All participants agreed that the tool enhances the analysis of QIMP algorithms.
This feedback reflects the situation, that apart from \qcvis{}, no participant has access to specialized visual tools for analyzing quantum images, only basic charts (see \autoref{sec:phases}).
P6 expressed reservations about the \modalityView{}, saying, "\textit{I'm not fully convinced that this visualization works. The use case only shows one example of the purple images changing to red. I want to test more examples, like coordinate flipping, to see if this visualization is useful.}"

\textbf{Usability and Interaction.}
Participants found the usability and interaction techniques of \qcvis{} intuitive and helpful. However, Q7, which assesses the ease of learning the system, received the lowest score in this category ($mean = 3.43$, $sd = 1.18$), indicating a moderate learning curve. P2 commented, "\textit{The system is not easy to learn, but that's to be expected since the entire application area is complicated.}" The mouse wheel interaction for switching color schemes in the \variationView{} was appreciated, with P4 suggesting, "\textit{This interaction would also be useful in the \modalityView{}.}"
Three participants noted issues with interaction clarity, specifically that it should be more visually obvious when the mouse hovers over a color scheme. P1 meant, "\textit{It is easy to accidentally trigger changes in views when simply scrolling through the image evolution.}"

\textbf{Visual Design.}
Participants gave positive feedback on the visual design. The \overview{} was particularly appreciated for its clarity. P4 remarked, "\textit{It is possible to read the intention of the gate to change a pixel, but through the image representations I can see at a glance that all pixels changed.}"
Participants P2-P5 preferred the \differenceView{} over the \absdifferenceView{}. Although its bars were smaller, it conveyed more helpful information for them.


\section{Discussion}
This section summarizes our findings from the design and evaluation of \qcvis{}, discusses its limitations, and future plans.

\subsection{General Findings}
Tailored visualization enables new insights. \qcvis{} enabled participants to gain new insights into the preparation and processing of NEQR images. As shown in \autoref{fig:UseCase1} (A2), the simulator exhibits minor inaccuracies that were previously unknown to the experts. Through the inspection of distributions in the \probabilityView{}, participants discovered that in a correctly implemented NEQR gate, the probability of a particular color is exactly $\frac{1}{m}$, where $m$ is the total number of pixels in the image. So the larger the images are, the lower the probability for the color of a specific pixel.

Complex application cases require simple visual designs. While developing \qcvis{}, we observed that highly complicated phenomena require cautious design of visualization techniques. During regular meetings with \expert{}, feedback showed that some earlier designs were too complicated, resulting in a steep learning curve. Simplifying these representations improved accessibility and ease of understanding.


\subsection{Limitations and Future Work}
\qcvis{} relies on simulated full statevectors, which are exponentially large and expensive to compute. While these are valuable to QIMP designers, it limits compatibility with real quantum hardware, which do not provide full statevector information. Extending \qcvis{} towards partial measurement data will enable support for real quantum device data.
We tested NEQR images up to a size of 8×8 pixels, which was appropriate for our experts. Future work needs to evaluate the tool's effectiveness with larger images to ensure scalability. In addition other quantum image schemes such as the Flexible Representation of Quantum Image (FRQI)~\cite{lang_representation_2024} could be explored. 
In our study, several features have been proposed to enhance \qcvis{}.
Support for 3D images is crucial to increase the applicability of \qcvis{} in the domain of XCT.
Another extension is the visualization of auxiliary qubits. Providing insight into the general states of these qubits would facilitate the investigation and debugging of gates that do not directly manipulate image structure, but instead perform other computational tasks.

\section{Conclusion}
Overall, \qcvis{} demonstrates a promising new approach for interactive visualization of QIMP algorithms using NEQR images. \qcvis{} integrates several visualization techniques to provide a comprehensive understanding of how quantum circuits prepare and process images, by allowing users to trace gate operations step-by-step, inspect pixel-wise distributions, and identify errors in circuit implementations. Our design study conducted with domain experts reveals, that \qcvis{} facilitates visual debugging and significantly improves the analysis of QIMP algorithms. Participants appreciated the system's detailed insights, although they experienced a considerable learning curve, and suggested enhancements such as 3D image support and auxiliary qubit visualization. These findings underscore the importance of tailored visualization tools in advancing QIMP research and provide a foundation for future improvements in this emerging field.


\acknowledgments{%
  We thank M. Blaimer, K. Dremel, M. Firsching, T. Fuchs, S. Kasperl, A. Papadaki, and D. Prjamkov for their feedback. This work was supported in part by the Bavarian Ministry of Economic Affairs, Regional Development and Energy within the Project "BayQS: Bayerisches Kompetenzzentrum Quanten Security and Data Science" under the number Az. 20-13-3410.3-2020.%
}

\bibliographystyle{abbrv-doi-hyperref}

\bibliography{template}

\appendix 

\end{document}